\documentclass[superscriptaddress,twocolumn,aps,showpacs,pra,preprintnumbers]{revtex4-1}
\usepackage{graphicx}
\usepackage{amsmath}

\usepackage{xcolor}     %% color options
\usepackage{soul}       %% underline, strike out
\begin{document}

\title{Isothermal compressibility determination across Bose-Einstein condensation}

\author{F. J. Poveda-Cuevas}\email{jacksonpc@ifsc.usp.br}
\affiliation{Instituto de F\'{i}sica de S\~{a}o Carlos, Universidade de S\~{a}o
Paulo, C.P. 369, 13560-970 S\~{a}o Carlos, SP, Brazil}
\author{P. C. M. Castilho}
\affiliation{Instituto de F\'{i}sica de S\~{a}o Carlos, Universidade de S\~{a}o
Paulo, C.P. 369, 13560-970 S\~{a}o Carlos, SP, Brazil}
\author{E. D. Mercado-Gutierrez}
\affiliation{Instituto de F\'{i}sica de S\~{a}o Carlos, Universidade de S\~{a}o
Paulo, C.P. 369, 13560-970 S\~{a}o Carlos, SP, Brazil}
\author{A. R. Fritsch}
\affiliation{Instituto de F\'{i}sica de S\~{a}o Carlos, Universidade de S\~{a}o
Paulo, C.P. 369, 13560-970 S\~{a}o Carlos, SP, Brazil}
\author{S. R. Muniz}
\affiliation{Instituto de F\'{i}sica de S\~{a}o Carlos, Universidade de S\~{a}o
Paulo, C.P. 369, 13560-970 S\~{a}o Carlos, SP, Brazil}
\author{E. Lucioni}
\affiliation{LENS and Dipartimento di Fisica e Astronomia, Universit\'a di Firenze, 50019 Sesto Fiorentino, Italy}
\author{G. Roati}
\affiliation{LENS and Dipartimento di Fisica e Astronomia, Universit\'a di Firenze, 50019 Sesto Fiorentino, Italy}
\author{V. S. Bagnato}
\affiliation{Instituto de F\'{i}sica de S\~{a}o Carlos, Universidade de S\~{a}o
Paulo, C.P. 369, 13560-970 S\~{a}o Carlos, SP, Brazil}

\date{\today}

\begin{abstract}
We apply the global thermodynamic variables approach to experimentally determine the isothermal compressibility parameter $\kappa_T$ of a trapped Bose gas across the phase transition. We demonstrate the behavior of $\kappa_T$ around the critical pressure, revealing the second order nature of the phase transition. Compressibility is the most important susceptibility to characterize the system. The use of global variables shows advantages with respect to the usual local density approximation method and can be applied to a broad range of situations.
\end{abstract}


\maketitle

\section{Introduction}
%% Paragraph 1 
It is well known that a gas composed of bosonic atoms with repulsive interparticle interaction at appropriate values of density and temperature undergoes Bose-Einstein condensation (BEC), a phase transition which shares similarities to transitions to superfluid and superconductive states. Since the first experimental demonstration of BEC \cite{Anderson-science269,Davis-prl75,Bradley-prl75}, efforts have directed toward in investigating the thermodynamic properties of such a macroscopic quantum system and finding suitable theoretical descriptions of the phase transitions \cite{Shi-physrep304}. Recently there is a revival of experimental interest devoted to the study of the thermodynamics of quantum gases. On one hand, distinguished works have explored the thermodynamics: a Fermi gas with repulsive interactions \cite{Lee-pra85}, a Fermi gas in the limit of very strong interactions, i.e., near the unitary regime \cite{Nascimbene-nature463,Ku-Science335}, a Fermi gas in a three-dimensional optical lattice showing fermionic Mott-insulator transition \cite{Duarte-prl114}, and the Boson gas in a two-dimensional optical lattice showing a bosonic Mott-Insulator transition \cite{Gemelke-nature460}. On the other hand, works on weakly interacting bosonic gases have demonstrated that, even in this simpler system, the understanding and characterization of the thermodynamic behavior, especially across the phase transition, are not yet complete \cite{Hung-nature470, Donner-science315, Olivares-Quiroz-jphysbatomphys43} and that more experimental works is needed to validate the theoretical predictions \cite{Goswami-jlowtempphys172, Floerchinger-pra79, Stanley-revmodphys71, Tarasov-pra90}. New approaches to investigating these systems and new experimental results can therefore contribute, in general, to advance the understanding of the thermodynamics of quantum gases and, in particular, of their phase transitions.

%% Paragraph 2
In this work, we experimentally determine a global susceptibility from a global thermodynamical variables approach for a harmonically trapped Bose gas \cite{Romero-Rochin-prl94,Romero-Rochin-bjp35,Sandoval-Figueroa-pre78}. We investigate and characterize the behavior of the susceptibility when the gas undergoes a BEC. In standard thermodynamics, the equivalent quantity of the global susceptibility that we define in this work is the isothermal compressibility. This parameter describes the relative variation of the volume $V$ of a system due to a change in the pressure $P$ at constant temperature $T$: $k_T=-\frac{1}{V}\left(\frac{\partial V}{\partial P}\right)_{N,T}$. It is a property associated with density fluctuations and it can also be expressed in terms of a second derivative of the free energy with respect to the pressure. At a second-order phase transition it is therefore expected to show a singularity. Here we provide experimental evidence of such a singular behavior by taking advantage of the global thermodynamic approach.

\section{Thermodynamics based on global variables}
%% Paragraph 3
Global variables have already been successfully employed to obtain the phase diagram \cite{Romero-Rochin-pra85} and measure the heat capacity \cite{Shiozaki-pra90} of a gas in a harmonic potential. The need to review standard thermodynamics when dealing with quantum gases comes naturally from the fact that they are usually trapped in nonhomogeneous (normally harmonic) potentials. In this situation standard definitions of pressure and volume do not apply. In fact, $P$ and $V$ are conjugate variables of thermodynamical systems defined for homogeneous densities. In particular, $P$ is an intensive variable having the same value in every position inside the volume occupied by the gas. The local density approximation (LDA) is often used in non-homogeneous situations to define local variables. A different approach, involving a set of thermodynamic variables with single values for the entire gas, allows a global description of the thermodynamics of an inhomogeneous gas and of its phase transitions. This global approach is particularly suited, compared to the LDA, for the case in which the gas is characterized by abrupt spatial variations of the density, as in the occurrence of a phase transition or in a more exotic situation such as in the presence of vortices or local potential impurities. 

%% Paragraph 4
The use of global variables to describe the thermodynamics of an inhomogeneous system has been extensively described elsewhere \cite{Romero-Rochin-prl94,Romero-Rochin-bjp35,Sandoval-Figueroa-pre78}. In brief, within the basis of thermodynamic and statistical mechanics one can infer a volume parameter and a pressure parameter respectively:
\begin{equation}
{\cal V}=\frac{1}{\omega_x\omega_y\omega_z}, \label{eq:V}
\end{equation}	
\begin{equation}
\Pi=\frac{2}{3{\cal V}}\langle U({\bf r})\rangle=\frac{m}{3{\cal V}}\int d^3r~n({\bf r})(\omega^2_xx^2+\omega^2_yy^2+\omega^2_zz^2),\label{eq:Pi}
\end{equation}
where $\omega_i$ with $\left(i=x,y,z\right)$ are the harmonic trap frequencies, $ \langle U({\bf r})\rangle $ is the spatial mean of the external potential, and $n({\bf r})$ is the density of the sample. ${\cal V}$ is a natural extensive ``volume'' for the trapped gas and the thermodynamic limit can be achieved by making the density parameter $n_{\cal V}=N/{\cal V}$ constant as $N$ and ${\cal V}$ grow indefinitely. $\Pi$ is its intensive conjugated variable $(\Pi=-\left(\frac{\partial F}{\partial{\cal V}}\right)_{N,T})$, where $F=F(N,{\cal V} ,T)$ is the Helmholtz free energy. A nice proof that $\Pi = \Pi\left( N,{\cal V},T \right)$ and ${\cal V}$ are a good set of variables to describe the system is obtained through the determination of the heat capacity, $C_{\cal V}$ \cite{Shiozaki-pra90}, whose behavior is close that one expected from treatment of a harmonic trapped Bose gas \cite{Grossmann-physletta208, Giorgini-jlowtempphys109}. In this framework, the isothermal compressibility parameter can be obtained from the following relation:
\begin{equation}
\kappa_T=-\frac{1}{{\cal V}}\left(\frac{\partial{\cal V}}{\partial\Pi}\right)_{N,T}.
\end{equation}
$\kappa_T$ is a quantity with the same properties of the standard compressibility $k_T$ \cite{Yukalov-pra72, Yukalov-laserphyslett2} and indicates the thermodynamic stability defined by the second derivative of Gibbs free energy. The convexity property of the free energy is maintained with the condition, $0 \leq \kappa_T < \infty$. Therefore, with this susceptibility we characterize a system in thermodynamic equilibrium \cite{Yukalov-laserphys23}.

\section{Experimental system and measurement}
%% Paragraph 5
We performed the measurements to determine $\kappa_T$ across the transition from a thermal cloud to a BEC of $^{87}{\rm Rb}$ atoms with a new experimental setup in which the volume parameter can be easily varied. The system is built in a standard double magneto-optical trap (MOT) configuration \cite{Myatt-oplett21}. In the first vacuum cell we load a MOT of $10^8$ atoms from a dispenser and then we transfer the atoms to the second cell using an on-resonance beam. Here, we recapture the atoms in a second MOT and, after performing a sub-Doppler cooling, we spin polarize the atomic sample in the hyperfine state $F=2,\, m_F=2$. Afterwards, we transfer the atoms at temperatures of about $40~\mu{\rm K}$, in a pure quadrupole magnetic trap where a first radio-frequency evaporation is performed. Simultaneously, we ramp on a far-detuned beam (with wavelength, $\lambda=1064~{\rm nm}$) focused on a waist $w_0=85~\mu{\rm m}$, dislocated by $z_0=300~\mu{\rm m}$ along the gravity direction below the center of the quadrupole trap. When the temperature of the atomic cloud decreases to approximately $10~\mu {\rm K}$, atoms migrate from the quadrupole trap to the center of the beam, which serves as an optical dipole trap (ODT). At that point we reduce the vertical magnetic-field gradient to a value that no longer compensates for the gravity. The atoms are thus confined in a hybrid trap given by the combination of the optical and magnetic confinements \cite{Lin-pra79}. Here we further decrease the temperature of the cloud by a second stage of radio-frequency evaporation followed by optical evaporation obtained by exponentially ramping down the power of the laser beam. We can eventually achieve a pure BEC of $\sim10^5$ atoms at typical temperatures of $100--200~{\rm nK}$. The hybrid potential, including gravity can be described by the following expression:
\begin{eqnarray}
U({\bf r})=\mu B'_x\sqrt{x^2+\frac{y^2}{2}+\frac{z^2}{2}}-\frac{U_0}{(1+y^2/y^2_R)}\nonumber\\
\exp\left[-\frac{2x^2+2(z-z_0)^2}{w_0^2(1+y^2/y^2_R)}\right]+mg(z-z_0)+E_0
\end{eqnarray}
$\mu$ is the atomic magnetic moment, $B'_x$ is the gradient of the quadrupole trap along the $x$ direction, $y_R = w^2_0\pi/\lambda$ is the Rayleigh range of the beam which propagates along direction $y$ and $U_0$ is the optical trap depth. $g$ is the gravitational acceleration, $m$ is the atomic mass and $E_0$ is the energy difference between the zero-field point absent the dipole trap and the total trap minimum, giving the trap minimum $U({\bf r}_{min})=0$ \cite{Lin-pra79}. At low temperatures the effective potential of the  HT can be safely approximated by a three dimensional harmonic potential, whose frequencies are	
\begin{equation}
\omega_x\simeq\omega_z=\sqrt{\frac{4U_0}{m w_0^2}},~~\omega_y=\sqrt{\frac{\mu B'_x}{2 m \left| z_0 \right|}}.\label{eq:freq}
\end{equation}
The trap has a cylindrical symmetry where the radial frequency confinement is due the ODT and the axial weaker confinement is due to magnetic-field gradient.

\begin{figure}[t!]
\includegraphics[width=1.0\columnwidth]{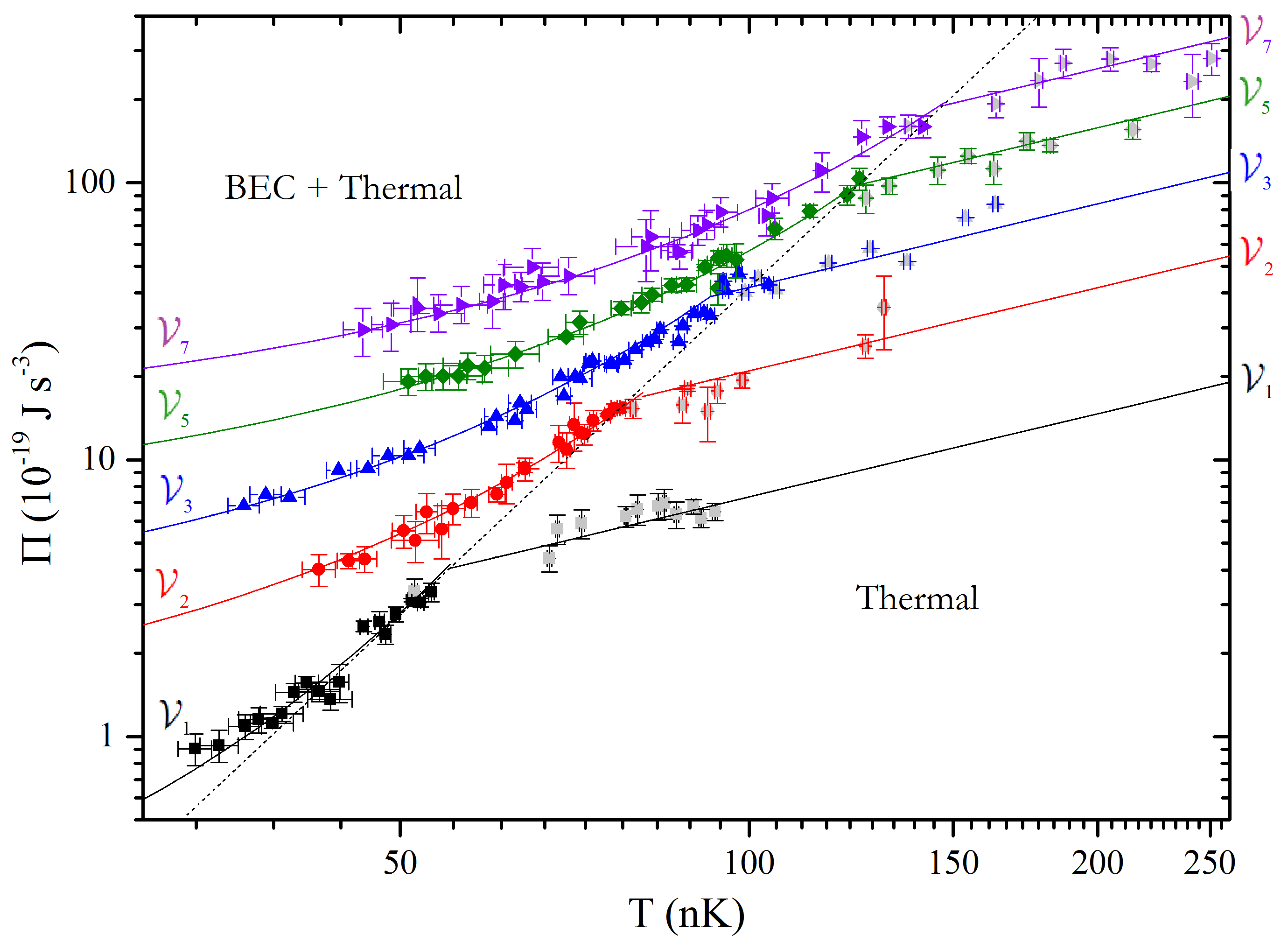}
\caption{(Color online) Pressure parameter vs temperature for a constant number of atoms ($N=1\times10^5$) and different volume parameters: ${\cal V}_{1}=1.9 \times 10^{-7} {\rm s}^3$, ${\cal V}_{2}=6.4\times 10^{-8}~{\rm s}^3$, ${\cal V}_{3}=3.2\times 10^{-8}~{\rm s}^3$, ${\cal V}_{4}=2.1\times 10^{-8}~{\rm s}^3$, ${\cal V}_{5}=1.75\times 10^{-8}~{\rm s}^3$, ${\cal V}_{6}=1.4\times 10^{-8}~{\rm s}^3$, and ${\cal V}_{7}=1\times 10^{-8}~{\rm s}^3$. Solid lines above $T_c$ represent the ideal gas law, whereas below $T_c$ are empirical exponential fits. The dotted black line marks the transition between the thermal and the condensed regimes. Error bars represent the statistical error on the average.}
\label{PixT}
\end{figure}

%% Paragraph 6
We characterize the atomic cloud by using absorption imaging after a free expansion from the trapping potential with a time of flight of $30~{\rm ms}$. Each image is fitted to a two-dimensional bimodal distribution composed of a Gaussian function and a Thomas-Fermi function, which are known to properly describe the thermal and the condensed component of the gas, respectively. The number of particles and temperature are obtained from the fitted images following conventional procedure. The volume parameter can be easily changed by varying the radial frequencies of the hybrid confinement, which directly depend on the final laser power of the ODT. We consider measurements for seven different sets of frequencies, i.e., for seven different volume parameters. Different temperatures have been obtained by changing the radio-frequency evaporation ramp; in this way the initial conditions for the optical evaporation change, allow us to achieve different final temperatures with the same trapping frequencies since the final power the ODT is the same. For each volume parameter we have performed many experimental runs for temperatures within the 	range $40--400~{\rm nK}$ and postselected atomic clouds containing $(1\pm 0.1)\times10^5$ atoms to be taken in consideration. In order to calculate the pressure parameter $\Pi$ by performing the integral in Eq. (\ref{eq:Pi}), it is necessary to reconstruct the density profile of the atoms in the trap, $n(\mathbf{r})$, from the measured profiles in the time of flight and the trap frequencies. Toward this aim, for the thermal component we can safely assume a free expansion, whereas for the interacting condensed component we apply the Castin-Dum procedure \cite{Castin-prl77,Kagan-pra55}. In Fig. \ref{PixT} we plot the calculated $\Pi(T)$ for each volume parameter. With a decrease in the temperature the atomic gas undergoes BEC: at high temperatures we observe a linear dependence of the pressure parameter on $T$ until an abrupt change takes place at a critical temperature $T_c$, and the decrease become faster than linear. Above $T_c$, experimental data are well reproduced by the the ideal gas law $\Pi {\cal V} = N k_B T$, plotted in Fig. \ref{PixT} for the known number of particles and the different volumes. Below $T_c$ we perform a proper empirical exponential fit which follows the behavior of the experimental points. These fitting functions, in principle, are not related to any theoretical model. For each volume parameter we can extract the critical pressure for condensation: lower volumes demand higher pressure to condense. The transition line from a thermal atomic cloud to a BEC in the $\Pi{\cal V}$-plane is shown in Fig. \ref{logVxPi} marking the separation between the white (thermal) and the gray (BEC) zone. 

\section{Isotherms and determination of compressibility parameter}
\begin{figure}[t!]
\includegraphics[width=1.0\columnwidth]{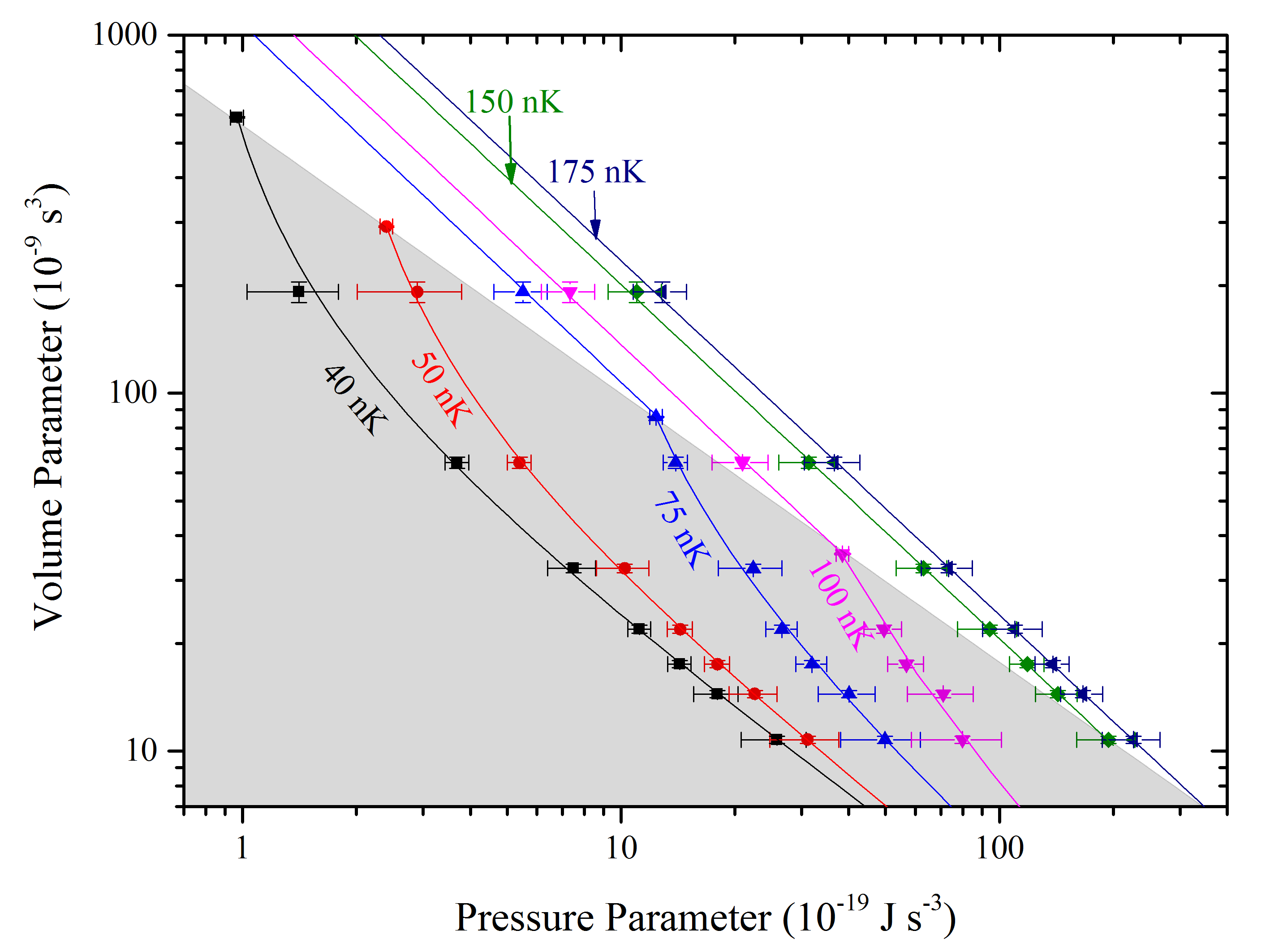}
\caption{(Color online) Isothermal ${\cal V}$ vs $\Pi$. Symbols represent the measured volume parameter vs the pressure parameter for different temperatures. The diagram shows the classical phase in white and the  quantum phase in gray, which are separated by the critical line on the ${\cal V} \Pi$ plane. In the thermal region the data obey the ideal gas and its behavior is linear in the log-log scale. On the other hand, in the BEC region the data exhibit a nonlinear behavior, indeed, we implement empirical fitting curves, known in the literature as the extended Langmuir adsorption isotherm equation, to follow each isotherm (these curves are no used in the analysis).  The error bars on the ${\cal V}$ axis come from error propagation of the measurement of the frequencies; on the other hand, the error bars on the $\Pi$ axis are associated with the exponential fit in Fig. \ref{PixT}.}
\label{logVxPi}
\end{figure}

%% Paragraph 7
From the measurements shown in Fig. \ref{PixT}, we extract different isotherms relating the volume and pressure parameters, ${\cal V}={\cal V}_T(\Pi)$, which we plot in Fig. \ref{logVxPi}. As the temperature decreases, the overall isothermal lines shift towards a lower pressure. We can clearly identify two different behaviors in the two different regions of the thermal and condensed regimes. In the thermal region, experimental points are well reproduced by the ideal gas law for the known number of atoms and temperatures (plotted as lines on the log-log scale of the figure). When an isotherm crosses the critical line for condensation an abrupt change occurs and it departs from the ideal gas behavior.

\begin{figure}[h!]
\includegraphics[width=1.0\columnwidth]{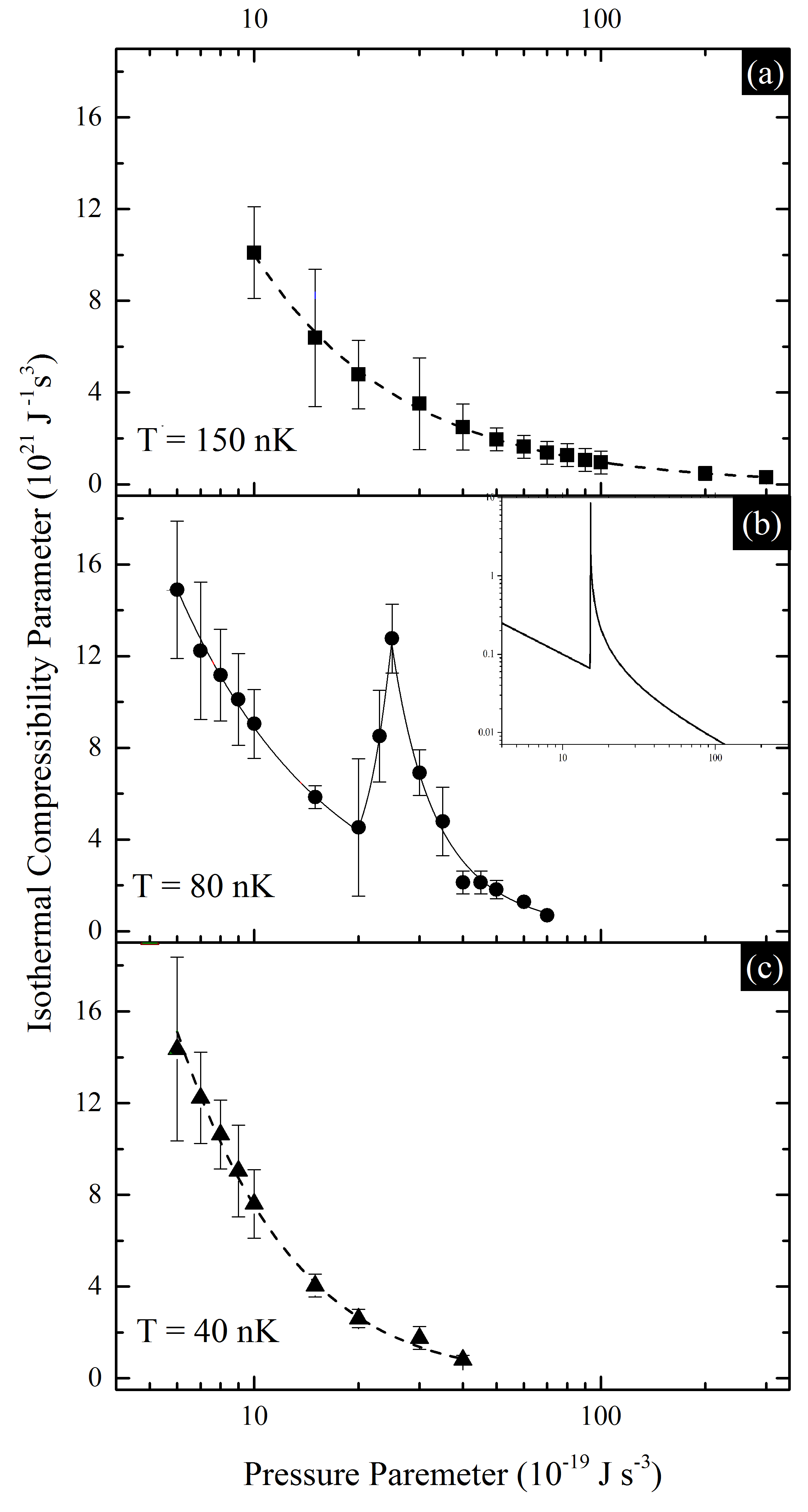}
\caption{Isothermal compressibility parameter vs pressure parameter for three temperatures: (a) $T=150~{\rm nK}$, (b) $T=80~{\rm nK}$, and (c) $T=40~{\rm nK}$. The inset in (b) is the $\kappa_T$ calculated from a simple toy model for the density distribution. Lines are guides for the eye. The error bar on the abscissa is not included so as to pollute the behavior of  $\kappa_T$. On the other hand, the error bar on the ordinate comes from the extrapolation of the tangent isothermal curve in Fig. \ref{logVxPi}}.
\label{kTxPi}
\end{figure}

%% Paragraph 8
We can now extract the isothermal compressibility $\kappa_T$ from derivation of the isotherms in Fig. \ref{logVxPi}. Derivation is performed point by point in correspondence with the experimental data in order not to rely on the arbitrary fitting curves, which do not correspond to any theoretical model. The obtained $\kappa_T$ values for three isothermal curves are shown in Fig. \ref{kTxPi}.  We have chosen the curves for $T=150~{\rm nK}$ [Fig. \ref{kTxPi}(a))], $T=80~{\rm nK}$ [Fig. \ref{kTxPi}(b)] and $T=40~{\rm nK}$ [Fig. \ref{kTxPi}(c)] because they demonstrate the three classes of behavior: pure thermal gas, gas undergoing BEC transition, and gas in the single BEC region, respectively. The isothermal curve at $150~{\rm nK}$ shows the decrease in $\kappa_T$ with $1/\Pi$, as expected for an ideal gas. Let us now consider the isotherm at $80~{\rm nK}$: at low pressures the gas is  thermal and the compressibility $\kappa_T$ decreases with increasing $\Pi$; when the pressure reaches the region between $20$ and $30~(\times 10^{-19}{\rm J}\cdot{\rm s}^{-3})$, the sudden increase in $\kappa_T$ indicates the transition. The compressibility reaches a maximum value before returning close to the base-line after $40 \times 10^{-19}{\rm J}\cdot{\rm s}^{-3}$. In this pressure range the compressibility acquires values 4 to 8 times higher than the base-line. The behavior of $\kappa_T$ in Fig. \ref{kTxPi} is typical for a second-order phase transition. An investigation of $\kappa_T$ vs $\Pi$ for different isothermal curves, where the transition takes place, reveals that at higher temperatures the transition occurs at a higher pressure and the peak of compressibility is broader for higher temperatures. Contrary to the expectation that quantities involving integration of density over the potential \cite{Ku-Science335} would be weakly sensitive to the phase transition, our data shows a sudden large variation in the compressibility at the thermal-BEC transition.

\section{Discussion}
%% Paragraph 9
We performed the data analysis using Castin-Dum procedure to reconstruct the \textit{in situ} density distribution starting with a Thomas-Fermi fit of the condensed component in the time-of-flight images. In order to probe that the general results we found do not depend on the specific model for the analysis, we also tested an alternative, less constrained, model. We fitted our images with two Gaussians for the thermal and condensed components and we reconstructed the \textit{in situ} profiles by applying a variational method \cite{Perez-Garcia-pra56, Teles-pra87, Teles-pra88} which has already proved to be valid to study the ballistic expansion dynamics of a condensate \cite{Teles-pra87, Teles-pra88}. We checked that the $\Pi(T)$ curves, and therefore all the derived thermodynamic quantities, extracted with the two reconstructing methods are quantitatively comparable.

%% Paragraph 10
A complete theory predicting the exact behavior of the compressibility parameter across the transition does not exist. Nevertheless, the need to make a prediction about the behavior and the shape of the compressibility around the critical point arises naturally. We have therefore attempted a comparison between our findings and the results of a toy model. We calculate $\Pi$ for synthetic density profiles consisting in a Gaussian thermal component and a Thomas-Fermi condensed one with a relative atom number given by the ideal BEC result. This model qualitatively catches the general experimental findings. In particular, the position and the shape of the compressibility peak are reproduced by the model as presented in the inset in Fig. \ref{kTxPi}. Nevertheless this simple model cannot give quantitative predictions, for example, of the absolute value of the compressibility, because it is over-simplified. A fair quantitative comparison would demand a more elaborate model, beyond the scope of this experimental report.

%% Paragraph 11
The introduction of the global variable approach has proven to be a valid complementary approach to the LDA. Generally speaking, the LDA approach in fact in fact has strong intrinsic limitations in the case where sudden variation of the densities occurs, as at the thermal-condensed interface in a Bose gas. In this situation the LDA would in fact require a very high imaging resolution, which is experimentally challenging. With the global approach we overcome this limitation by describing the system undergoing phase transition as a whole and we can provide evidence of the compressibility peak at the transition. On the other hand, the global variables approach needs many measurements for different volumes with the same atom number to trace a single isothermal curve, and this can be experimentally nontrivial. In this sense the LDA has the advantage of leading to a complete isothermal curve from the analysis of a single image. Due to the lack of experimental points, we cannot precisely measure the compressibility in the close vicinity of the phase transition. Nevertheless, the expected sharp peak in $\kappa_T$ near the critical point is quite clear and shares remarkable similarities to the behavior of the isothermal compressibility for liquid helium as observed across the $\lambda$-point \cite{Boghosian-pr152,Grilly-pr149,Elwell-pr164}. 

\section{Conclusions}
%% Paragraph 13
In this article, we have used the concept of global thermodynamic variables to measure, the most appropriate susceptibility to understand the phase transition, the isothermal compressibility parameter of a harmonically confined Bose gas. Once the sample had undergone BEC we characterized this phase transition, from the classical to the quantum regime, indicating a second-order transition likely related to a spontaneous symmetry breaking. The concept of using global variables to determine the global compressibility is quite useful in situations where the LDA cannot be applied. In another more complex physical systems in which there are abrupt changes in the density are of interest for superfluid physics, such as vortices, vortex lattices, solitons, \textit{inter alia}, and, especially, superfluid turbulence, recently demonstrated by our group \cite{Henn-prl103}. In this case the local variables do not make sense and the global behavior in the compressibility may indicate new characteristics of the turbulent regime. Such an investigation is currently in progress. 

We acknowledge financial support from FAPESP (Brazil), CNPq (Brazil), CAPES (Brazil), and LENS (Italy).

\bibliography{bibkT}
\end{document}